\documentstyle[11pt]{article}
\begin{document}

\title{{{\bf Reconstructing $f(R)$ modified gravity from ordinary and entropy-corrected versions of
the holographic and new agegraphic dark energy models}}}
\author{K. Karami$^{1,2}$\thanks{E-mail: KKarami@uok.ac.ir} , M.S. Khaledian$^{1}$\thanks{E-mail: MS.Khaledian@uok.ac.ir}\\
$^{1}$\small{Department of Physics, University
of Kurdistan, Pasdaran St., Sanandaj, Iran}\\$^{2}$\small{Research
Institute for Astronomy
$\&$ Astrophysics of Maragha (RIAAM), Maragha, Iran}\\
}

\maketitle

\begin{abstract}
Here, we peruse cosmological usage of the most promising candidates
of dark energy in the framework of $f(R)$ theory. We reconstruct the
different $f(R)$ modified gravity models in the spatially flat FRW
universe according to the ordinary and entropy-corrected versions of
the holographic and new agegraphic dark energy models, which
describe accelerated expansion of the universe. We also obtain the
equation of state parameter of the corresponding $f(R)$-gravity
models. We conclude that the holographic and new agegraphic
$f(R)$-gravity models can behave like phantom or quintessence
models. Whereas the equation of state parameter of the
entropy-corrected models can transit from quintessence state to
phantom regime as indicated by recent observations.
\end{abstract}
\clearpage
\section{Introduction}
Astrophysical data from type Ia supernovae, cosmic microwave
background radiation and large scale structure have provided
convincing evidence for the present observable universe to be in the
phase of accelerated expansion \cite{Riess}. One explanation for the
cosmic acceleration is the dark energy (DE), an exotic energy with
negative pressure. The origin and nature of DE is still a mystery. A
great variety of DE models have been proposed (for review see
\cite{Padmanabhan,Copeland}), but most of them are not able to
explain all features of the universe, or are artificially
constructed in the sense that it introduces too many free parameters
to be able to fit with the experimental data.

The holographic DE (HDE) is one of interesting DE candidates which
was proposed based on the holographic principle
\cite{Horava,Hooft,Fischler}. It was shown in \cite{cohen} that in
quantum field theory, the UV cut-off $\Lambda$ should be related to
the IR cut-off $L$ due to limit set by forming a black hole.
Following this line, Li \cite{Li} argued that the total energy in a
region of size $L$ should not exceed the mass of a black hole of the
same size, thus $L^3\rho_{\Lambda}\leq LM_P^2$, where
$\rho_{\Lambda}$ is the quantum zero point energy density caused by
UV cut-off $\Lambda$ and $M_P$ is the reduced Planck Mass
$M_P^{-2}=8\pi G$. Also Li \cite{Li} showed that the cosmic
coincidence problem can be resolved by inflation in the HDE model,
provided the minimal number of e-foldings \cite{Li}. The HDE models
have been studied widely in the literature
\cite{Enqvist,Elizalde2,Guberina1,Guberina2,Karami1}. Obviously, in
the derivation of HDE, the black hole entropy $S_{\rm BH}$ plays an
important role. As is well known, usually, $S_{\rm BH} = A/(4G)$,
where $A\sim L^2$ is the area of horizon. However, in the
literature, this entropy-area relation can be modified to
\cite{modak}
\begin{equation}
S_{\rm
BH}=\frac{A}{4G}+\tilde{\alpha}\ln{\frac{A}{4G}}+\tilde{\beta},\label{MEAR}
\end{equation}
where $\tilde{\alpha}$ and $\tilde{\beta}$ are dimensionless
constants of order unity. These corrections can appear in the black
hole entropy in loop quantum gravity (LQG) \cite{HW}. They can also
be due to thermal equilibrium fluctuation, quantum fluctuation, or
mass and charge fluctuations (for a good review see \cite{HW} and
references therein). More recently, motivated by the corrected
entropy-area relation (\ref{MEAR}) in the setup of LQG, the energy
density of the entropy-corrected HDE (ECHDE) was proposed by Wei
\cite{HW}.

Recently, the original agegraphic dark energy (ADE) model was
proposed by Cai \cite{Cai}. The ADE model assumes that the observed
DE comes from the spacetime and matter field fluctuations in the
universe \cite{Cai}. Following the line of quantum fluctuations of
spacetime, Karolyhazy et al. \cite{Kar1} discussed that the distance
$t$ in Minkowski spacetime cannot be known to a better accuracy than
$\delta{t}\sim t_{P}^{2/3}t^{1/3}$ where $t_P$ is the reduced Planck
time. Based on Karolyhazy relation, Maziashvili \cite{Maz} discussed
that the energy density of metric fluctuations of the Minkowski
spacetime is given by $\rho_{\Lambda}\sim 1/(t_P^2t^2)\sim
M_P^2/t^2$. Based on Karolyhazy relation \cite{Kar1} and Maziashvili
arguments \cite{Maz}, Cai proposed the original ADE model to explain
the accelerated expansion of the universe \cite{Cai}. Since the
original ADE model suffers from the difficulty to describe the
matter-dominated epoch, the new ADE (NADE) model was proposed by Wei
$\&$ Cai \cite {Wei1}, while the time scale was chosen to be the
conformal time instead of the age of the universe. It was found that
the coincidence problem could be solved naturally in the NADE model
\cite{Wei3}. The ADE models have given rise to a lot of interest
recently and have been examined and studied in ample detail in
\cite{Wei2,Kim,Kim11,Sheykhi}. More recently, very similar to the
ECHDE model, the energy density of the entropy-corrected NADE
(ECNADE) was proposed by Wei \cite{HW} and investigated in ample
detail by \cite{Karami2}.

Another explanation for the cosmic acceleration and what we get from
observational data is the different approach, so called ``modified
gravity''. This scheme possesses the relevant feature that
experimental data turn out to be naturally interpreted without the
need of additional components like DE (for review see
\cite{Capozziello}). In the situation when Einstein general
relativity (GR) cannot naturally describe the DE epoch of the
universe the search of alternative, modified gravity which is
consistent with solar system tests/observational data is of primary
interest \cite{NojiriOdin}. A very popular modified gravity model is
the so-called $f(R)$-gravity \cite{husawiski,noj abdalla,Noj2} where
$f(R)$ is an arbitrary function of the scalar curvature $R$.
Actually, among extended theories of modified gravity,
$f(R)$-gravity represents a viable alternative to DE and naturally
gives rise to accelerating singularity-free solution in early and
late cosmic epochs \cite{Noj}. For instance, the form $f(R)=R+R^2
+1/R$ can predict the unification of the early-time inflation and
late-time cosmic acceleration in the standard metric formulation
\cite{Noj4}.

Viewing the modified $f(R)$-gravity model as an effective
description of the underlying theory of DE, and considering the
ordinary and entropy-corrected versions of the HDE and NADE
scenarios as pointing in the direction of the underlying theory of
DE, it is interesting to study how the $f(R)$-gravity can describe
the HDE, ECHDE, NADE and ECNADE densities as effective theories of
DE models. This motivated us to establish the different models of
$f(R)$-gravity according to the ordinary and entropy-corrected
versions of the HDE and NADE scenarios. This paper is organized as
follows. In section 2, we review the theory of $f(R)$-gravity in the
metric formalism. In sections 3 to 7 we reconstruct the different
$f(R)$-gravity models corresponding to the HDE, ECHDE, NADE and
ECNADE models, respectively. Section 8 is devoted to our
conclusions.

\section{The theory of $f(R)$ modified gravity}

The general $f(R)$-gravity action is given by
\cite{NojiriOdin,husawiski}
\begin{equation}
S = \int\sqrt{-g}~{\rm d}^4 x\left[\frac{{R+f(R)}}{{2k^2 }} + L_{\rm
matter} \right],\label{action}
\end{equation}
where $k^2=M_P^{-2}=8\pi G$ and $\hbar=c=1$. Also $G$, $g$, $R$ and
$L_{\rm matter}$ are the gravitational constant, the determinant of
metric $g_{\mu\nu}$, the Ricci scalar and the lagrangian density of
the matter inside the universe, respectively.

One notes that there are in fact two strategies in $f(R)$ theories:
the metric formalism, where the action is varied with respect to the
metric only; and the Palatini formalism \cite{Ferraris}, where the
metric and the connection are treated as two independent variables
with respect to which the action is varied \cite{Wu}. It is only in
Einstein gravity $f(R)=0$ that both approaches reach the same
result. In general $f(R)$ theories, the problem of which approach
should be used is still an open question and the final solution may
be determined by further observations and theoretical development
\cite{Wu}. Here like \cite{Wu} we use the metric formalism.

Taking the variation of the action (\ref{action}) with respect to
the metric $g_{\mu\nu}$, one can obtain the field equations as
\cite{NojiriOdin,husawiski}
\begin{equation}
R_{\mu\nu}-\frac{1}{2}Rg_{\mu\nu}=
k^2\Big(T_{\mu\nu}^{(R)}+T_{\mu\nu}^{(m)}\Big),\label{field
equation}
\end{equation}
where
\begin{equation}
k^2T_{\mu\nu}^{(R)}=\frac{1}{2}g_{\mu\nu}f(R)-R_{\mu\nu}f'(R)+\big(\nabla_{\mu}\nabla_{\nu}-g_{\mu\nu}\Box\big)f'(R).
\end{equation}
Here $R_{\mu\nu}$ and $T_{\mu\nu}^{(m)}$ are the Ricci tensor and
the energy-momentum tensor of the matter, respectively. Also the
prime denotes a derivative with respect to $R$.

Now if we consider the spatially flat FRW metric for the universe as
\begin{equation}
{\rm d}s^2  =  - {\rm d}t^2  + a^2 (t)\sum\limits_{i = 1}^3 {({\rm
d}x^i )^2 },\label{FRW}
\end{equation}
and taking $T_\nu^{\mu(m)}={\rm diag}(-\rho_m,p_m,p_m,p_m)$ for the
energy-momentum tensor of the matter in the prefect fluid form, then
the set of field equations (\ref{field equation}) reduce to the
modified Friedmann equations in the framework of $f(R)$-gravity as
\begin{equation}
\frac{3}{k^2}H^2= \rho_m+\rho_R,\label{FiEq1}
\end{equation}
\begin{equation}
\frac{1}{k^2}\big(2\dot{H}+3H^2\big)=-(p_m+p_R),\label{FiEq2}
\end{equation}
where
\begin{equation}
\rho_{R} = \frac{1}{k^2}\left[-\frac{1}{2}f(R)+3\big(\dot{H}+
H^2\big)f'(R)-18\big(4 H^2\dot{H}+
H\ddot{H}\big)f''(R)\right],\label{density}
\end{equation}
\begin{eqnarray}
p_{R}= \frac{1}{k^2}\Big[\frac{1}{2}f(R)-\big(\dot{H}+3H^2\big)f'(R)
~~~~~~~~~~~~~~~~~~~~~~~~~~~~~~~~~~~~~~~~~~~~~~~~~\nonumber\\
+6\big(8H^2\dot{H}+6H\ddot{H}+4{\dot{H}}^2+\dot{\ddot{H}}\big)f''(R)+36
\big(\ddot{H}+4H\dot{H}\big)^2f'''(R)\Big],\label{pressure}
\end{eqnarray}
and
\begin{equation} R =6\big(\dot{H}+2H^2\big).\label{R}
\end{equation}
Here $H=\dot{a}/a$ is the Hubble parameter and the dot denotes a
derivative with respect to cosmic time $t$. Also $\rho_{R}$ and
$p_{R}$ are the curvature contribution to the energy density and
pressure.

The energy conservation laws are still given by
\begin{equation}
\dot{\rho_m}+3H(\rho_m+p_m)=0,
\end{equation}
\begin{equation}
\dot{\rho}_{R}+3H(\rho_{R}+p_{R})=0.\label{ecT}
\end{equation}
In the case of $f(R)=0$, from Eqs. (\ref{density}) and
(\ref{pressure}) we have $\rho_R=0$ and $p_R=0$. Therefore Eqs.
(\ref{FiEq1}) and (\ref{FiEq2}) transform to the usual Friedmann
equations in GR.

The equation of state (EoS) parameter due to the curvature
contribution is defined as \cite{Nozari}
\begin{eqnarray}
\omega_{R}=\frac{p_R}{\rho_R}~~~~~~~~~~~~~~~~~~~~~~~~~~~~~~~~~~~~~~~~~~~~~~~~~~~~~~~~~~~~~~~~~~~~~~~~~~~~~~~~~~~~
~~~~~~~~~~~~~~~~~~~\nonumber\\=-1-
\frac{4\Big[\dot{H}f'(R)+3\big(3H\ddot{H}-4H^2\dot{H}+4{\dot{H}}^2+\dot{\ddot{H}}\big)f''(R)
+18\big(\ddot{H}+4H\dot{H}\big)^2f'''(R)\Big]}
{\Big[f(R)-6\big(\dot{H}+ H^2\big)f'(R)+36\big(4 H^2\dot{H}+
H\ddot{H}\big)f''(R)\Big]}.~\label{wHDETotal}
\end{eqnarray}
Note that for a $f(R)$ dominated universe, Eq. (\ref{FiEq1}) yields
\begin{equation}
\frac{3}{k^2}H^2=\rho_R.
\end{equation}
Taking time derivative of the above relation and using the
continuity equation (\ref{ecT}), one can get the EoS parameter as
\begin{equation}
\omega_{R}=-1-\frac{2\dot{H}}{3H^2},
\end{equation}
which shows that for the phantom, $\omega_{R}<-1$, and quintessence,
$\omega_{R}>-1$, dominated universe, we need to have $\dot{H}>0$ and
$\dot{H}<0$, respectively.

For a given $a=a(t)$, by the help of Eqs. (\ref{density}) and
(\ref{pressure}) one can reconstruct the $f(R)$-gravity according to
any DE model given by the EoS $p_R=p_R(\rho_R)$ or
$\rho_R=\rho_R(a)$. There are two classes of scale factors which
usually people consider them for describing the accelerating
universe in $f(R)$, $f(\mathcal{G})$ and $f(R,\mathcal{G})$ modified
gravities \cite{Nojiri}.

The first class of scale factor is given by \cite{Nojiri,Sadjadi}
\begin{equation}
a(t)=a_0(t_s-t)^{-h},~~~t\leq t_s,~~~h>0.\label{a}
\end{equation}
Using Eqs. (\ref{R}) and (\ref{a}) one can obtain
\begin{equation}
H=\frac{h}{t_s-t}=\left[\frac{h}{6(2h+1)}R\right]^{1/2},~~~\dot{H}=H^2/h,\label{respect
to r}
\end{equation}
which $\dot{H}=H^2/h>0$ shows that the model (\ref{a}) is
correspondence to a phantom dominated universe. This is why that in
the literature the model (\ref{a}) is usually so-called the phantom
scale factor.

For the second class of scale factor defined as \cite{Nojiri}
\begin{equation}
a(t)=a_0t^h,~~~h>0,\label{aQ}
\end{equation}
one can obtain
\begin{equation}
H=\frac{h}{t}=\left[\frac{h}{6(2h-1)}R\right]^{1/2},~~~\dot{H}=-H^2/h,\label{respect
to rQ}
\end{equation}
which $\dot{H}=-H^2/h<0$ reveals that the model (\ref{aQ}) describes
a quintessence dominated universe. Hence, the model (\ref{aQ}) is
so-called the quintessence scale factor in the literature.

In sections 3 to 6 using the two classes of scale factors (\ref{a})
and (\ref{aQ}), we reconstruct the different $f(R)$-gravities
according to the HDE, ECHDE, NADE and ECNADE models.

\section{$f(R)$ reconstruction from HDE model}

Here we reconstruct the $f(R)$-gravity according to the HDE
scenario. Following Li \cite{Li} the HDE density in a spatially flat
universe is given by
\begin{equation}
\rho_{\Lambda}=\frac{3c^2}{k^2R_h^2},\label{ro H}
\end{equation}
where $c$ is a numerical constant. Recent observational data, which
have been used to constrain the HDE model, show that for the flat
universe $c=0.818_{-0.097}^{+0.113}$ \cite {Li6}. Also $R_h$ is the
future event horizon defined as
\begin{equation}
R_h=a\int_t^{\infty}\frac{{\rm d}t}{a}=a\int_a^{\infty}\frac{{\rm
d}a}{Ha^2}.\label{L0}
\end{equation}
For the first class of scale factor (\ref{a}) and using Eq.
(\ref{respect to r}), the future event horizon $R_h$ yields
\begin{equation}
R_h=a\int_t^{t_s}\frac{{\rm
d}t}{a}=\frac{t_s-t}{h+1}=\frac{1}{h+1}\sqrt{\frac{6h(2h+1)}{R}}.\label{L}
\end{equation}
Replacing Eq. (\ref{L}) into (\ref{ro H}) one can get
\begin{equation}
\rho_\Lambda=\frac{c^2(h+1)^2}{2k^2h(2h+1)}R.\label{ro H R}
\end{equation}
Substituting Eq. (\ref{ro H R}) in the differential equation
(\ref{density}), i.e. $\rho_R=\rho_{\Lambda}$, gives the following
solution
\begin{equation}
f(R)=\lambda_+ R^{m_+}+\lambda_- R^{m_-}+\gamma_c~R,\label{frHDE}
\end{equation}
where
\begin{equation}
m_\pm=\frac{3+h\pm\sqrt{h^2-10h+1}}{4},\label{alpha}
\end{equation}
and
\begin{equation}
\gamma_c=-\frac{c^2(h+1)^2}{h^2}.\label{gammac}
\end{equation}
Also $\lambda_\pm$ are the integration constants that can be
determined from the necessary boundary conditions. Following
\cite{NojiriOdin1} the accelerating expansion in the present
universe could be generated, if one consider that $f(R)$ could be a
small constant at present universe, that is
\begin{equation}
f(R_0)=-2R_0,\label{bc1}
\end{equation}
\begin{equation}
f'(R_0)\sim0,\label{bc2}
\end{equation}
where $R_0\sim(10^{-33}{\rm eV})^2$ is the current curvature.
Applying the above boundary conditions to the solution (\ref{frHDE})
one can obtain
\begin{equation}
\lambda_{+}=\frac{\gamma_c (m_--1)+2m_-}{(m_+-m_-)R_0^{m_+-1}},
\end{equation}
\begin{equation}
\lambda_{-}=\frac{\gamma_c (m_+-1)+2m_+}{(m_--m_+)R_0^{m_--1}}.
\end{equation}
Replacing Eq. (\ref{frHDE}) into (\ref{wHDETotal}) and using
(\ref{respect to r}) one can get the EoS parameter of the
holographic $f(R)$-gravity model as
\begin{equation} \omega_{R}=-1-\frac{2}{3h},~~~h>0,\label{wHDE2}
\end{equation}
which corresponds to a phantom accelerating universe, i.e.
$\omega_R<-1$. Recent observational data indicates that the EoS
parameter $\omega_{R}$ at the present lies in a narrow strip around
$\omega_{R} = -1$ and is quite consistent with being below this
value \cite{Copeland}.

For the second class of scale factor (\ref{aQ}), using Eq.
(\ref{respect to rQ}) the future event horizon $R_h$ reduces to
\begin{equation}
R_h=a\int_t^{\infty}\frac{{\rm
d}t}{a}=\frac{t}{h-1}=\frac{1}{h-1}\sqrt{\frac{6h(2h-1)}{R}},~~~h>1,\label{RhQ}
\end{equation}
where the condition $h>1$ is obtained due to have a finite future
event horizon. If we repeat the above calculations then one can
obtain the both of $f(R)$ and $\omega_R$ corresponding to the HDE
for the second class of scale factor (\ref{aQ}). The result for
$f(R)$ is same as (\ref{frHDE}) where now
\begin{equation}
m_\pm=\frac{3-h\pm\sqrt{h^2+10h+1}}{4},\label{alphaQ}
\end{equation}
\begin{equation}
\gamma_c=-\frac{c^2(h-1)^2}{h^2}.\label{gammacQ}
\end{equation} Also the EoS parameter is obtained as
\begin{equation}
\omega_{R}=-1+\frac{2}{3h},~~~h>1,\label{wHDEQ}
\end{equation}
which describes an accelerating universe with the quintessence EoS
parameter, i.e. $-1<\omega_R<-1/3$.
\section{$f(R)$ reconstruction from ECHDE model}

Using the corrected entropy-area relation (\ref{MEAR}), the energy
density of the ECHDE can be obtained as \cite{HW}
\begin{equation}
\rho_\Lambda=\frac{3c^2}{k^2R_h^2}+\frac{\alpha}{R_h^4}\ln\left(\frac{R_h^2}{k^2}\right)+\frac{\beta}{R_h^4},\label{ECHDE}
\end{equation}
where $\alpha$ and $\beta$ are dimensionless constants of order
unity. In the special case $\alpha=\beta=0$, the above equation
yields the well-known HDE density (\ref{ro H}). Since the last two
terms in Eq. (\ref{ECHDE}) can be comparable to the first term only
when $R_h$ is very small, the corrections make sense only at the
early stage of the universe. When the universe becomes large, ECHDE
reduces to the ordinary HDE \cite{HW}.

For the first class of scale factor (\ref{a}), substituting Eq.
(\ref{L}) into (\ref{ECHDE}) yields
\begin{equation}
\rho_\Lambda=\frac{c^2(h+1)^2}{2k^2h(2h+1)}R+\frac{(h+1)^4}{36h^2(2h+1)^2}
\left[\alpha\ln\left(\frac{6h(2h+1)}{k^2(h+1)^2R}\right)+\beta\right]R^2.\label{ECHDE
R}
\end{equation}
Solving the differential equation (\ref{density}) for the energy
density (\ref{ECHDE R}) gives the entropy-corrected holographic
$f(R)$-gravity as
\begin{eqnarray}
f(R)=\lambda_+R^{m_+}+\lambda_-R^{m_-}+\gamma_cR
~~~~~~~~~~~~~~~~~~~~~~~~~~~~~~~~~~~~~~~~~~~~~~~~~~~~~~~~~~~~
\nonumber\\+\frac{k^2(h+1)^4}{54h^2(2h+1)}\left\{\alpha\left[\left(\frac{h-5}{3}\right)-
\ln\left(\frac{6h(2h+1)}{k^2(h+1)^2R}\right)\right]-\beta\right\}R^2,\label{frECHDE}
\end{eqnarray}
where $m_{\pm}$ and $\gamma_c$ are given by Eqs. (\ref{alpha}) and
(\ref{gammac}), respectively. Also $\lambda_\pm$ are determined from
the boundary conditions (\ref{bc1}) and (\ref{bc2}) as
\begin{eqnarray}
\lambda_+=\frac{\gamma_c (m_--1)+2m_-}{(m_+-m_-)R_0^{m_+-1}}
+\frac{k^2(h+1)^4(m_--2)}{54h^2(2h+1)(m_+-m_-)R_0^{m_+-2}}\times~~~~~~~~~~~~~~~~~
\nonumber\\
\left\{\alpha\left[\frac{h-5}{3}-\frac{1}{m_-{-2}}-\ln\left({\frac{6h(2h+1)}
{k^2(h+1)^2R_0}}\right)\right]-\beta\right\} ,\label{c1}
\end{eqnarray}
\begin{eqnarray}
\lambda_-=\frac{\gamma_c (m_+-1)+2m_+}{(m_--m_+)R_0^{m_--1}}
+\frac{k^2(h+1)^4(m_+-2)}{54h^2(2h+1)(m_--m_+)R_0^{m_--2}}\times~~~~~~~~~~~~~~~~~
\nonumber\\
\left\{\alpha\left[\frac{h-5}{3}-\frac{1}{m_+{-2}}-\ln\left({\frac{6h(2h+1)}
{k^2(h+1)^2R_0}}\right)\right]-\beta\right\}.\label{c2}
\end{eqnarray}
Substituting Eq. (\ref{frECHDE}) into (\ref{wHDETotal}) one can get
the EoS parameter of the entropy-corrected holographic
$f(R)$-gravity model as
\begin{equation}
\omega_{R}=-1-\frac{2}
{3h}\left\{1+\frac{k^2(h+1)^2\Big[-\alpha+\alpha\ln\Big({\frac{6h(2h+1)}
{k^2(h+1)^2R}}\Big)+\beta\Big]R}{18c^2h(2h+1)+k^2(h+1)^2\Big[\alpha\ln\Big({\frac{6h(2h+1)}
{k^2(h+1)^2R}}\Big)+\beta\Big]R}\right\},~~~h>0,\label{wHDE3}
\end{equation}
which can be rewritten using the first relation of Eq. (\ref{respect
to r}) as
\begin{equation}
\omega_{R}=-1-\frac{2}
{3h}\left\{1+\frac{-\alpha+2\alpha\ln\Big({\frac{h}
{k(h+1)H}}\Big)+\beta}
{3c^2\Big(\frac{h}{k(h+1)H}\Big)^2+2\alpha\ln\Big(\frac{h}
{k(h+1)H}\Big)+\beta}\right\},~~~h>0.\label{wECHDE}
\end{equation}
The above relation shows that the EoS parameter is time-dependent
and in contrast with constant EoS parameter (\ref{wHDE2}), it can
justify the transition from quintessence state, $\omega_R>-1$, to
the phantom regime, $\omega_R<-1$, as indicated by recent
observations \cite{Sahni}. Note that if we set $\alpha=\beta=0$ then
Eqs. (\ref{frECHDE}) and (\ref{wECHDE}) reduce to (\ref{frHDE}) and
(\ref{wHDE2}), respectively.

For the second class of scale factor (\ref{aQ}), the result of
$f(R)$ is obtained as
\begin{eqnarray}
f(R)=\lambda_+R^{m_+}+\lambda_-R^{m_-}+\gamma_cR
~~~~~~~~~~~~~~~~~~~~~~~~~~~~~~~~~~~~~~~~~~~~~~~~~
\nonumber\\+\frac{k^2(h-1)^4}{54h^2(2h-1)}\left\{\alpha\left[\left(\frac{h+5}{3}\right)+
\ln\left(\frac{6h(2h-1)}{k^2(h-1)^2R}\right)\right]+\beta\right\}R^2,\label{frECHDEQ}
\end{eqnarray}
where
\begin{eqnarray}
\lambda_+=\frac{\gamma_c (m_--1)+2m_-}{(m_+-m_-)R_0^{m_+-1}}
+\frac{k^2(h-1)^4(m_--2)}{54h^2(2h-1)(m_+-m_-)R_0^{m_+-2}}\times~~~~~~~~~~~~~~~~~
\nonumber\\
\left\{\alpha\left[\frac{h+5}{3}+\frac{1}{m_-{-2}}+\ln\left({\frac{6h(2h-1)}
{k^2(h-1)^2R_0}}\right)\right]+\beta\right\} ,\label{c1}
\end{eqnarray}
\begin{eqnarray}
\lambda_-=\frac{\gamma_c (m_+-1)+2m_+}{(m_--m_+)R_0^{m_--1}}
+\frac{k^2(h-1)^4(m_+-2)}{54h^2(2h-1)(m_--m_+)R_0^{m_--2}}\times~~~~~~~~~~~~~~~~~
\nonumber\\
\left\{\alpha\left[\frac{h+5}{3}+\frac{1}{m_+{-2}}+\ln\left({\frac{6h(2h-1)}
{k^2(h-1)^2R_0}}\right)\right]+\beta\right\},\label{c2}
\end{eqnarray}
and the parameters $m_{\pm}$ and $\gamma_c$ are given by Eqs.
(\ref{alphaQ}) and (\ref{gammacQ}), respectively. Also the EoS
parameter is obtained as
\begin{equation}
\omega_{R}=-1+\frac{2}
{3h}\left\{1+\frac{-\alpha+2\alpha\ln\Big({\frac{h}
{k(h-1)H}}\Big)+\beta}
{3c^2\Big(\frac{h}{k(h-1)H}\Big)^2+2\alpha\ln\Big(\frac{h}
{k(h-1)H}\Big)+\beta}\right\},~~~h>1.\label{wECHDEQ}
\end{equation}
Here also to have a finite $R_h$, the parameter $h$ should be in the
range of $h>1$. One notes that the EoS parameter (\ref{wECHDEQ}) is
also dynamical and in contrast with constant EoS parameter
(\ref{wHDEQ}), it can accommodate the transition from $\omega_R>-1$
to $\omega_R<-1$ at recent stage.

\section{$f(R)$ reconstruction from NADE model}

Following \cite{Wei1}, the energy density of the NADE is given by
\begin{equation}
\rho_{\Lambda}=\frac{3{n}^2}{k^2\eta^2},\label{NADE}
\end{equation}
where the numerical factor 3$n^2$ is introduced to parameterize some
uncertainties, such as the species of quantum fields in the
universe, the effect of curved spacetime (since the energy density
is derived for Minkowski spacetime), and so on. It was found that
the coincidence problem could be solved naturally in the NADE model
provided that the single model parameter $n$ is of order unity
\cite{Wei3}. The joint analysis of the astronomical data for the
NADE model in flat universe gives the best-fit value (with 1$\sigma$
uncertainty) $n=2.716_{-0.109}^{+0.111}$ \cite{Wei3}. Also $\eta$ is
the conformal time of FRW universe, and given by
\begin{equation}
\eta=\int\frac{{\rm d}t}{a}=\int\frac{{\rm d}a}{Ha^2}.\label{eta}
\end{equation}
For the first class of scale factor (\ref{a}), the conformal time
$\eta$ by the help of Eq. (\ref{respect to r}) yields
\begin{equation}
\eta=\int_t^{t_s}\frac{{\rm
d}t}{a}=\frac{(t_s-t)^{h+1}}{a_0(h+1)}=\frac{1}{a_0(h+1)}\left[\frac{6h(2h+1)}{R}\right]^{\frac{h+1}{2}}.\label{E}
\end{equation}
Substituting Eq. (\ref{E}) into (\ref{NADE}) one can obtain
\begin{equation}
\rho_\Lambda=\frac{3n^2a_0^2(h+1)^2}{k^2\big(6h(2h+1)\big)^{h+1}}R^{h+1}.\label{ro
NEDE R}
\end{equation}
Solving the differential equation (\ref{density}) for the energy
density (\ref{ro NEDE R}) reduces to
\begin{equation}
f(R)=\lambda_+R^{m_+}+\lambda_-R^{m_-}+
\gamma_n~R^{h+1},\label{frNADE}
\end{equation}
where
\begin{equation}
\gamma_n=-\frac{n^2a_0^2(h+1)^2}{h^2(h+2)\big(6h(2h+1)\big)^h},\label{gamman}
\end{equation}
and $m_{\pm}$ are given by Eq. (\ref{alpha}).  Using the boundary
conditions (\ref{bc1}) and (\ref{bc2}) the parameters
$\lambda_{\pm}$ are determined as
\begin{equation}
\lambda_{+}=\frac{\gamma_n\big(m_--(h+1)\big)R_0^{h}+2m_-
}{(m_+-m_-)R_0^{m_+-1}},
\end{equation}
\begin{equation}
\lambda_{-}=\frac{\gamma_n\big(m_+-(h+1)\big)R_0^{h}+2m_+
}{(m_--m_+)R_0^{m_--1}}.
\end{equation}
Replacing Eq. (\ref{frNADE}) into (\ref{wHDETotal}) one can get the
EoS parameter of the new agegraphic $f(R)$-gravity model as
\begin{equation}
\omega_{R}=-1-\frac{2(h+1)}{3h},~~~h>0,\label{wNADE}
\end{equation}
which like the EoS parameter of the holographic $f(R)$-gravity model
(\ref{wHDE2}), it always crosses the phantom-divide line, i.e.
$\omega_R<-1$.

For the second class of scale factor (\ref{aQ}), using (\ref{respect
to rQ}) the conformal time $\eta$ is obtained as
\begin{equation}
\eta=\int_0^t\frac{{\rm
d}t}{a}=\frac{t^{1-h}}{a_0(1-h)}=\frac{1}{a_0(1-h)}\left[\frac{6h(2h-1)}{R}\right]^{\frac{1-h}{2}},
~~~\frac{1}{2}<h<1,\label{EQ}
\end{equation}
where the condition $\frac{1}{2}<h<1$ is necessary due to have a
real finite conformal time. The result of $f(R)$ is
\begin{equation}
f(R)=\lambda_+R^{m_+}+\lambda_-R^{m_-}+
\gamma_n~R^{1-h},\label{frNADEQ}
\end{equation}
where
\begin{equation}
\lambda_{+}=\frac{\gamma_n\big(m_-+h-1\big)R_0^{-h}+2m_-
}{(m_+-m_-)R_0^{m_+-1}},
\end{equation}
\begin{equation}
\lambda_{-}=\frac{\gamma_n\big(m_++h-1\big)R_0^{-h}+2m_+
}{(m_--m_+)R_0^{m_--1}},
\end{equation}
with
\begin{equation}
\gamma_n=\frac{n^2a_0^2(h-1)^2}{h^2(h-2)\big(6h(2h-1)\big)^{-h}},\label{gammanQ}
\end{equation}
and the parameters $m_{\pm}$ are given by Eq. (\ref{alphaQ}).

Also the EoS parameter is obtained as
\begin{equation}
\omega_{R}=-1+\frac{2(1-h)}{3h},~~~\frac{1}{2}<h<1,\label{wNADEQ}
\end{equation}
which shows a quintessence-like EoS parameter with
$-1<\omega_R<-1/3$.
\section{$f(R)$ reconstruction from ECNADE model}

With the help of quantum corrections to the entropy-area relation
(\ref{MEAR}) in the setup of LQG, the energy density of the ECNADE
is given by \cite{HW}
\begin{eqnarray}
\rho_{\Lambda} = \frac{3n^2}{k^2\eta^2} +
\frac{\alpha}{\eta^4}\ln{\left(\frac{\eta^2}{k^2}\right)} +
\frac{\beta}{\eta^4},\label{ECNADE}
\end{eqnarray}
which closely mimics to that of ECHDE density (\ref{ECHDE}) and
$R_h$ is replaced with the conformal time $\eta$. Here $\alpha$ and
$\beta$ are dimensionless constants of order unity. In the special
case $\alpha=\beta=0$, the above equation yields the well-known NADE
density (\ref{NADE}).

For the first class of scale factor (\ref{a}), substituting Eq.
(\ref{E}) into (\ref{ECNADE}) one can get
\begin{eqnarray}
\rho_\Lambda=\frac{3n^2a_0^2(h+1)^2}{k^2\big(6h(2h+1)\big)^{h+1}}R^{h+1}
~~~~~~~~~~~~~~~~~~~~~~~~~~~~~~~~~~~~~~~~~~~~~~~
\nonumber\\+\frac{a_0^4(h+1)^4}{\big(6h(2h+1)\big)^{2h+2}}
\left[\alpha\ln\left(\frac{\big(6h(2h+1)\big)^{h+1}}
{k^2a_0^2(h+1)^2R^{h+1}}\right)+\beta\right]R^{2h+2}.\label{ro
ECNEDE R}
\end{eqnarray}
Solving the differential equation (\ref{density}) for the energy
density (\ref{ro ECNEDE R}) yields
\begin{eqnarray}
f(R)=\lambda_+R^{m_+}+\lambda_-R^{m_-}+\gamma_n~R^{h+1}
-\frac{k^2a_0^4(h+1)^4}{3h(3+10h+6h^2)\big(6h(2h+1)\big)^{2h+1}}\times~~~~
\nonumber\\
\left\{\alpha\left[\frac{(7h+5)(h+1)}{3+10h+6h^2}+
\ln\left(\frac{\big(6h(2h+1)\big)^{h+1}}{k^2a_0^2(h+1)^2R^{h+1}}\right)\right]+\beta\right\}R^{2h+2},~~
\label{frECNADE}
\end{eqnarray}
where $m_{\pm}$ and $\gamma_n$ are given by Eqs. (\ref{alpha}) and
(\ref{gamman}), respectively. Also $\lambda_\pm$ are determined from
the boundary conditions (\ref{bc1}) and (\ref{bc2}) as
\begin{eqnarray}
\lambda_+=\frac{\gamma_n\big(m_--(h+1)\big)R_0^{h}+2m_-
}{(m_+-m_-)R_0^{m_+-1}}~~~~~~~~~~~~~~~~~~~~~~~~~~~~~~~~~~~~~~~~~~~~~~~~~~~~~~~~~
\nonumber\\+\frac{k^2a_0^4(2h+2-m_-)(h+1)^4R_0^{2h+2-m_+}}
{3h(3+10h+6h^2)\big(6h(2h+1)\big)^{2h+1}(m_+-m_-)}\times~~~~~~~~~~~~~~~~~~~~~~~~~~~~
\nonumber\\
\left\{\alpha\left[\frac{(7h+5)(h+1)}{3+10h+6h^2}+\frac{h+1}{m_-{-2}-2h}
+\ln\left({\frac{\big(6h(2h+1)\big)^{h+1}}{k^2a_0^2(h+1)^2R_0^{h+1}}}\right)\right]+\beta\right\},\label{c3}
\end{eqnarray}
\begin{eqnarray}
\lambda_-=\frac{\gamma_n\big(m_+-(h+1)\big)R_0^{h}+2m_+
}{(m_--m_+)R_0^{m_--1}}~~~~~~~~~~~~~~~~~~~~~~~~~~~~~~~~~~~~~~~~~~~~~~~~~~~~~~~~~
\nonumber\\+\frac{k^2a_0^4(2h+2-m_+)(h+1)^4R_0^{2h+2-m_-}}
{3h(3+10h+6h^2)\big(6h(2h+1)\big)^{2h+1}(m_--m_+)}\times~~~~~~~~~~~~~~~~~~~~~~~~~~~~
\nonumber\\
\left\{\alpha\left[\frac{(7h+5)(h+1)}{3+10h+6h^2}+\frac{h+1}{m_+{-2}-2h}
+\ln\left({\frac{\big(6h(2h+1)\big)^{h+1}}{k^2a_0^2(h+1)^2R_0^{h+1}}}\right)\right]+\beta\right\}.\label{c4}
\end{eqnarray}
Replacing Eq. (\ref{frECNADE}) into (\ref{wHDETotal}) yields the EoS
parameter of the entropy-corrected new agegraphic $f(R)$-gravity
model as
\begin{equation}
\omega_{R}=-1-\frac{2(h+1)}{3h} \left\{1+
\frac{-\alpha+\alpha\ln\left({\frac{\big(6h(2h+1)\big)^{h+1}}{k^2a_0^2(h+1)^2R^{h+1}}}\right)
+\beta}
{\frac{3n^2}{k^2a_0^2(h+1)^2}\big(\frac{6h(2h+1)}{R}\big)^{h+1}+
\alpha\ln\left(\frac{\big(6h(2h+1)\big)^{h+1}}{k^2a_0^2(h+1)^2R^{h+1}}\right)+\beta}
\right\},~h>0.\label{wECNADE1}
\end{equation}
It can be rewritten using the first relation of Eq. (\ref{respect to
r}) as
\begin{equation}
\omega_{R}=-1-\frac{2(h+1)}{3h}
\left\{1+\frac{-\alpha+2\alpha\ln\Big(\frac{1}{ka_0(h+1)}\big(\frac{h}{H}\big)^{h+1}\Big)+\beta}
{3n^2\Big(\frac{1}{ka_0(h+1)}\big(\frac{h}{H}\big)^{h+1}\Big)^2+2\alpha\ln\Big(\frac{1}{ka_0(h+1)}\big(\frac{h}{H}\big)^{h+1}\Big)+\beta}
\right\},~h>0,\label{wECNADE2}
\end{equation}
which is time-dependent and in contrast with constant EoS parameter
(\ref{wNADE}), it can justify the transition from $\omega_R>-1$ to
$\omega_R<-1$. Note that if we set $\alpha=\beta=0$ then Eqs.
(\ref{frECNADE}) and (\ref{wECNADE2}) reduce to (\ref{frNADE}) and
(\ref{wNADE}), respectively.

For the second class of scale factor (\ref{aQ}), the result of
$f(R)$ is obtained as
\begin{eqnarray}
f(R)=\lambda_+R^{m_+}+\lambda_-R^{m_-}+\gamma_n~R^{1-h}
+\frac{k^2a_0^4(h-1)^4}{3h(3-10h+6h^2)\big(6h(2h-1)\big)^{1-2h}}\times~~
\nonumber\\
\left\{\alpha\left[\frac{(7h-5)(h-1)}{3-10h+6h^2}+
\ln\left(\frac{\big(6h(2h-1)\big)^{1-h}}{k^2a_0^2(h-1)^2R^{1-h}}\right)\right]+\beta\right\}R^{2-2h},~~
\label{frECNADEQ}
\end{eqnarray}
where
\begin{eqnarray}
\lambda_+=\frac{\gamma_n\big(m_-+(h-1)\big)R_0^{-h}+2m_-
}{(m_+-m_-)R_0^{m_+-1}}~~~~~~~~~~~~~~~~~~~~~~~~~~~~~~~~~~~~~~~~~~~~~~~~~~~~~~~~~
\nonumber\\-\frac{k^2a_0^4(2-2h-m_-)(h-1)^4R_0^{2-2h-m_+}}
{3h(3-10h+6h^2)\big(6h(2h-1)\big)^{1-2h}(m_+-m_-)}\times~~~~~~~~~~~~~~~~~~~~~~~~~~~~
\nonumber\\
\left\{\alpha\left[\frac{(7h-5)(h-1)}{3-10h+6h^2}+\frac{1-h}{m_-{-2}+2h}
+\ln\left({\frac{\big(6h(2h-1)\big)^{1-h}}{k^2a_0^2(h-1)^2R_0^{1-h}}}\right)\right]+\beta\right\},\label{c3}
\end{eqnarray}
\begin{eqnarray}
\lambda_-=\frac{\gamma_n\big(m_++(h-1)\big)R_0^{-h}+2m_+
}{(m_--m_+)R_0^{m_--1}}~~~~~~~~~~~~~~~~~~~~~~~~~~~~~~~~~~~~~~~~~~~~~~~~~~~~~~~~~
\nonumber\\-\frac{k^2a_0^4(2-2h-m_+)(h-1)^4R_0^{2-2h-m_-}}
{3h(3-10h+6h^2)\big(6h(2h-1)\big)^{1-2h}(m_--m_+)}\times~~~~~~~~~~~~~~~~~~~~~~~~~~~~
\nonumber\\
\left\{\alpha\left[\frac{(7h-5)(h-1)}{3-10h+6h^2}+\frac{1-h}{m_+{-2}+2h}
+\ln\left({\frac{\big(6h(2h-1)\big)^{1-h}}{k^2a_0^2(h-1)^2R_0^{1-h}}}\right)\right]+\beta\right\},\label{c4}
\end{eqnarray}
and the parameters $m_{\pm}$ and $\gamma_n$ are given by Eqs.
(\ref{alphaQ}) and (\ref{gammanQ}), respectively. Also the EoS
parameter is obtained as
\begin{equation}
\omega_{R}=-1+\frac{2(1-h)}{3h}
\left\{1+\frac{-\alpha+2\alpha\ln\Big(\frac{1}{ka_0(1-h)}\big(\frac{h}{H}\big)^{1-h}\Big)+\beta}
{3n^2\Big(\frac{1}{ka_0(1-h)}\big(\frac{h}{H}\big)^{1-h}\Big)^2+2\alpha\ln\Big(\frac{1}{ka_0(1-h)}
\big(\frac{h}{H}\big)^{1-h}\Big)+\beta}
\right\},~\frac{1}{2}<h<1.\label{wECNADEQ}
\end{equation}
Here also to have a real finite conformal time $\eta$, the parameter
$h$ should be in the range of $\frac{1}{2}<h<1$. Contrary to the
constant EoS parameter (\ref{wNADEQ}), the dynamical EoS parameter
(\ref{wECNADEQ}) can accommodate the transition from $\omega_R>-1$
to $\omega_R<-1$ at recent stage.
\section{$f(R)$ reconstruction in de Sitter space}
The scale factor in de Sitter space is defined as
\begin{equation}
a(t)=a_0e^{Ht},~~~H={\rm constant},\label{adS}
\end{equation}
which can describe the early-time inflation of the universe
\cite{Nojiri}. Using Eqs. (\ref{R}) and (\ref{adS}) one can obtain
\begin{equation}
H=\left(\frac{R}{12}\right)^{1/2}.\label{respect to rD}
\end{equation}
Then Eq. (\ref{density}) takes the form
\begin{equation}
4k^2\rho_{R} = -2f(R)+Rf'(R).\label{densityD}
\end{equation}
Also the EoS parameter (\ref{wHDETotal}) yields $\omega_R=-1$ which
behaves like the cosmological constant.

For the scale factor (\ref{adS}), using Eq. (\ref{respect to rD})
the future event horizon $R_h$ reduces to
\begin{equation}
R_h=a\int_t^{\infty}\frac{{\rm
d}t}{a}=H^{-1}=\left(\frac{R}{12}\right)^{-1/2}.\label{RhQ}
\end{equation}
Substituting Eq. (\ref{RhQ}) into HDE density (\ref{ro H}) one can
get
\begin{equation}
\rho_{\Lambda}=\frac{c^2}{4k^2}R.\label{ro H D}
\end{equation}
Replacing Eq. (\ref{ro H D}) in the differential equation
(\ref{densityD}), i.e. $\rho_R=\rho_{\Lambda}$, gives the solution
\begin{equation}
f(R)=\lambda R^{2}-c^2R,\label{frHDE D}
\end{equation}
where $\lambda$ is an integration constant. Note that in order to
generate the inflation at the early universe as in Starobinsky's
model \cite{Nojiri,Starobinsky}, one may require
\begin{equation}
\lim_{R\rightarrow\infty} f(R)\propto R^2.\label{inf}
\end{equation}
We see that the holographic $f(R)$-gravity model (\ref{frHDE D}) can
satisfy the requirement of having to inflation (\ref{inf}).

Replacing Eq. (\ref{RhQ}) into ECHDE density (\ref{ECHDE}) one can
obtain
\begin{equation}
\rho_{\Lambda}=\frac{c^2}{4k^2}R+\left(\frac{R}{12}\right)^2\left[\beta+\alpha\ln\left(\frac{12}{k^2R}\right)\right].\label{ro
ECH D}
\end{equation}
Solving the differential equation (\ref{densityD}) for the energy
density (\ref{ro ECH D}) yields the entropy-corrected holographic
$f(R)$-gravity model as
\begin{equation}
f(R)=\lambda R^{2}-c^2R+\left(\frac{kR}{6}\right)^2
\left\{\beta\ln{(R)}-\frac{\alpha}{2}\left[\ln{\left(\frac{12}{k^2R}\right)}\right]^2\right\},\label{frECHDE
D}
\end{equation}
which in the absence of correction terms, i.e. $\alpha=\beta=0$,
recovers the result of holographic $f(R)$-gravity model (\ref{frHDE
D}). As we already mentioned the correction terms in ECHDE density
(\ref{ro ECH D}) become important in the early inflation era.
Equation (\ref{frECHDE D}) also confirms that besides the term
$\lambda R^2$, the corrections make sense during the inflation when
$R\rightarrow\infty$.

The conformal time $\eta$ for the scale factor (\ref{adS}) yields
\begin{equation}
\eta=\int_0^t\frac{{\rm
d}t}{a}=\frac{1}{a_0H}\Big(1-e^{-Ht}\Big).\label{etadS1}
\end{equation}
Here to obtain $\eta=\eta(R)$ one cannot replace $t$ by $R$ in
(\ref{etadS1}). Therefore for the scale factor (\ref{adS}) one
cannot obtain the $f(R)$-gravity models corresponding to the NADE
(\ref{NADE}) and ECNADE (\ref{ECNADE}) densities. To avoid of this
problem we set $t\rightarrow\infty$ for the upper limit of the
integral (\ref{etadS1}). Hence the result yields
\begin{equation}
\eta=\int_0^{\infty}\frac{{\rm
d}t}{a}=(a_0H)^{-1}=\left(\frac{a_0^2R}{12}\right)^{-1/2}.\label{etadS}
\end{equation}
Replacing Eq. (\ref{etadS}) into NADE density (\ref{NADE}) yields
\begin{equation}
\rho_{\Lambda}=\frac{n^2a_0^2}{4k^2}R.\label{ro NADE D}
\end{equation}
Substituting Eq. (\ref{ro NADE D}) in the differential equation
(\ref{densityD}) gives the solution
\begin{equation}
f(R)=\lambda R^{2}-n^2a_0^2R,\label{frNADE D}
\end{equation}
where $\lambda$ is an integration constant. Note that the new
agegraphic $f(R)$-gravity model (\ref{frNADE D}) like (\ref{frHDE
D}) satisfies the inflation condition (\ref{inf}).

Replacing Eq. (\ref{etadS}) into ECNADE density (\ref{ECNADE}) one
can get
\begin{equation}
\rho_{\Lambda}=\frac{n^2a_0^2}{4k^2}R+\left(\frac{a_0^2R}{12}\right)^2\left[\beta+\alpha\ln\left(\frac{12}{k^2a_0^2R}\right)\right].\label{ro
ECN D}
\end{equation}
Solving the differential equation (\ref{densityD}) for the energy
density (\ref{ro ECN D}) yields the entropy-corrected new agegraphic
$f(R)$-gravity model as
\begin{equation}
f(R)=\lambda R^{2}-n^2a_0^2R+\left(\frac{ka_0^2R}{6}\right)^2
\left\{\beta\ln{(R)}-\frac{\alpha}{2}\left[\ln{\left(\frac{12}{k^2a_0^2R}\right)}\right]^2\right\},\label{frECNADE
D}
\end{equation}
which for $\alpha=\beta=0$, the above result reduces to the new
agegraphic $f(R)$-gravity model (\ref{frNADE D}). Here also like
(\ref{frECHDE D}), during the inflation era not only the term
$\lambda R^2$ but also the correction terms become considerable in
Eq. (\ref{frECNADE D}).

\section{Conclusions}

Here, we considered the ordinary and entropy-corrected versions of
the HDE and NADE models which are originated from some significant
features of quantum gravity. The HDE is motivated from the
holographic hypothesis \cite{Horava,Hooft,Fischler} and the NADE is
originated form uncertainty relation of quantum mechanics together
with the gravitational effect in GR \cite{Kar1,Maz}. Among various
candidates to explain cosmic accelerated expansion, only HDE and
NADE models are based on the entropy-area relation. However, this
definition can be modified from the inclusion of quantum effects,
motivated from the LQG. Hence the ECHDE and ECNADE were introduced
by addition of correction terms to the energy densities of HDE and
NADE, respectively \cite{HW}.


We investigated the HDE, ECHDE, NADE and ECNADE in the framework of
$f(R)$-gravity. Among other approaches related with a variety of DE
models, a very promising approach to DE is related with the modified
theories of gravity known as $f(R)$-gravity, in which DE emerges
from the modification of geometry. Modified gravity gives a natural
unification of the early-time inflation and late-time acceleration
thanks to different role of gravitational terms relevant at small
and at large curvature and may naturally describe the transition
from deceleration to acceleration in the cosmological dynamics
\cite{Granda}. We reconstructed the different theories of modified
gravity based on the $f(R)$ action in the spatially flat FRW
universe for the three classes of scale factors containing i)
$a=a_0(t_s-t)^{-h}$, ii) $a=a_0t^h$ and iii) $a=a_0e^{Ht}$ and
according to the original and entropy-corrected versions of the HDE
and NADE scenarios. Furthermore, we obtained the EoS parameters of
the corresponding $f(R)$-gravity models. Our calculations show that
for the first class of scale factor, the EoS parameter of the
holographic and new agegraphic $f(R)$-gravity models always crosses
the phantom-divide line. Whereas for the second class, the EoS
parameter of the mentioned models behaves like the quintessence EoS
parameter. The EoS parameter of the entropy-corrected holographic
and new agegraphic $f(R)$-gravity models for the both of first and
second classes of scale factors can accommodate the transition from
quintessence state, $\omega_R>-1$, to the phantom regime,
$\omega_R<-1$, as indicated by recent observations. For the third
scale factor, the EoS parameter behaves like the cosmological
constant. Also the $f(R)$-gravity models corresponding to the HDE,
ECHDE, NADE and ECNADE can predict the early-time inflation of the
universe.

Note that although $f(R)$ theories offer a chance to explain the
acceleration of the universe, they are not free of problems. As an
alternative to DE for driving the late-time cosmic acceleration, the
$f(R)$-gravity needs to pass the cosmological tests, including the
constraints about the cosmic expansion and the cosmic structure
formation. In addition, as a modified gravity theory, it needs to
pass the solar system test, such as the constraints on the
Brans-Dicke theory \cite{Henttunen}. Besides, in many models of
$f(R)$-gravity, including the ones called realistic, the
``fine-tuning'' and the ``cosmic coincidence'' problems which are
the two well-known difficulties of the cosmological constant
problems, have not been essentially solved. Although there remain
possibilities to solve the coincidence problem (see e.g.
\cite{Bisabr}), the existence of small parameters in $f(R)$-gravity
models is still the most important problem. In case of $\Lambda$CDM
model, the scale of the cosmological constant is very small compared
with the Planck scale, which is unnatural, especially from the
viewpoint of the unified theory of the particle physics. Even in the
realistic $f(R)$ models like Hu-Sawicki's one \cite{Sawicki} which
satisfies both cosmological and solar system tests, there appear the
small parameters, which would be still unnatural.
\\
\\
\noindent{{\bf Acknowledgements}}\\
The authors thank the anonymous referee for a number of valuable
suggestions. The authors also thank Professor Shin'ichi Nojiri for
useful discussions. The work of K. Karami has been supported
financially by Research Institute for Astronomy and Astrophysics of
Maragha, Iran.

\end{document}